\documentstyle[12pt,epsf]{article}

\makeatletter
\@addtoreset{equation}{section}

\newcommand{\nn}{\nonumber}
\newcommand{\vs}[1]{\vspace*{#1}}
\newcommand{\hs}[1]{\hspace*{#1}}
\newcommand{\tr}{\mathop{\rm Tr}}
\newcommand{\p}{\partial}
\newcommand{\Half}{\frac12}
\newcommand{\unit}{\hbox to 3.8pt{\hskip1.3pt \vrule height 7.4pt
    width .4pt \hskip.7pt \vrule height 7.85pt width .4pt \kern-2.4pt 
    \hrulefill \kern-3pt \raise 3.7pt\hbox{\char'40}}}

\def\href#1#2{#2}  
\newcommand{\vv}{\mbox{\boldmath$v$}}
\newcommand{\rr}{\mbox{\boldmath$r$}}
\newcommand{\hx}{\widehat{x}}
\newcommand{\PB}[2]{\{#1,#2\}}
\newcommand{\de}{\delta_{\epsilon}}

\begin{document}

\vspace{10mm}
\begin{titlepage}
\title{
\hfill\parbox{4cm}
{\normalsize KUNS-1612\\{\tt hep-th/9910196}}\\
\vspace{1cm}
Brane Configuration from Monopole Solution\\
in Non-Commutative Super Yang-Mills Theory
}
\author{
Koji {\sc Hashimoto}\thanks{{\tt hasshan@gauge.scphys.kyoto-u.ac.jp}},
{}
Hiroyuki {\sc Hata}\thanks{{\tt hata@gauge.scphys.kyoto-u.ac.jp}}
{} and
Sanefumi {\sc Moriyama}\thanks{{\tt moriyama@gauge.scphys.kyoto-u.ac.jp}}
\\[7pt]
{\it Department of Physics, Kyoto University, Kyoto 606-8502, Japan}
}
\date{\normalsize October, 1999}
\maketitle
\thispagestyle{empty}

\vs{10mm}

\begin{abstract}
\normalsize\noindent 
We study the structure of the monopole configuration in $U(2)$
non-commutative super Yang-Mills theory.
Our analysis consists of two steps: solving the BPS equation and then
the eigenvalue equation in the non-commutative space.
Calculation to the first non-trivial order in the non-commutativity
parameter $\theta$ shows that the monopole exhibits a certain
non-locality. This structure is precisely the one expected from the  
recent predictions by the brane-configuration technique.

\end{abstract}

\end{titlepage}


\section{Introduction}

The fertile developments in string theory in this half a decade
have enabled us to understand various perturbative and non-perturbative 
phenomena in field theories by geometrical intuitive pictures. This
progress in string theory is now beyond the province of reproduction
of the result of the field theories, that is, now the string theory
has predictive power in various field theories.

One of the most intriguing examples is the 1/4 BPS dyon solution in
4-dimensional ${\cal N}\!=\!4$ super Yang-Mills theory (SYM). The
study of the $1/4$ BPS states in SYM was triggered by the
discovery of  the stable string network in type IIB
string theory \cite{Schwarz:1996,Dasgupta:1997pu}. When this string
network has its ends on D3-branes, these states preserve 1/4
supersymmetries of the original D3-brane worldvolume theory
\cite{Bergman:1997yw}. After the study from the string theory side,
there appeared some works
\cite{Hashimoto:1998zs,Hashimoto:1998nj,Kawano:1998bp,Lee:1998nv} 
in which explicit field theoretical solutions for the corresponding
solitons were constructed. The properties of the solution favor the
string theory interpretation with respect to their $(p,q)$-charges,
masses and supersymmetries. 

\begin{figure}[tdp]
\begin{center}
\parbox[b]{130mm}{
\begin{center}
\leavevmode
\epsfxsize=80mm
\put(200,35){D3}
\put(200,115){D3}
\put(50,55){D-string}
\put(110,-8){$\delta$}
\epsfbox{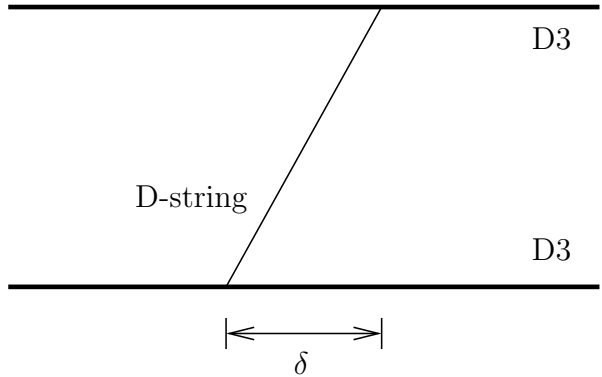}
\caption{The D-string suspended between two parallel D3-branes tilts
  owing to the background of the constant NS-NS $B$-field.}
\label{fig1}
\end{center}
}
\end{center}
\end{figure}

Recent topics of the string-theoretical realization of field theories
are concerning the space non-commutativity
\cite{Connes:1998cr,Douglas:1998fm,Seiberg:1999vs}. The
non-commutative super Yang-Mills theory (NCSYM) is realized as a
decoupling limit of the worldvolume theory on D3-branes in the
non-trivial NS-NS 2-form background. Taking advantage of this
equivalence, some basic properties of localized objects in this exotic 
field theory were analyzed \cite{Hashimoto:1999}. The brane
configurations corresponding to the monopoles, dyons and 1/4 BPS dyons
were constructed, and they were shown to have the same masses and 
supersymmetry properties as the ordinary SYM counterparts. One
fascinating prediction from this brane technique is that the monopole
in the NCSYM has non-locality $\delta$ due to the tilt of the D-string
suspended between two D3-branes (see fig.\ \ref{fig1}). Believing that
the brane configuration of this figure precisely captures the field
theoretical properties, the configuration of the monopole in the NCSYM
should reproduce the tilted line, as the eigenvalues of the Higgs
field. The ends of the D-string appear to be magnetic charges, hence
the field theoretical solution should contain dipole structure.

In this paper, we explicitly solve the BPS equation for the monopole
of the NCSYM  to the first non-trivial order in $\theta_{ij}$ which
specifies the non-commutativity. This 1/2 BPS solution has the same
mass as the ordinary SYM monopole, in agreement with the prediction by
string theory. Solving the non-commutative eigenvalue equation in
Sec.\ 3, we show in Sec.\ 4 that the solution actually reproduces the
tilt of the suspended D-string. Examining the magnetic field, the
dipole structure is also found. In the final section, we summarize the
paper and give some discussions.

\section{Non-commutative BPS equation and its solution}

The Bogomol'nyi-Prasad-Sommerfield (BPS) monopole solution 
\cite{Bogomol'nyi:1976,Prasad:1975kr} of 
the ordinary SYM is saturating a particular energy bound which is
usually called the BPS bound. Since this bound is topologically
sensible, the state saturating the bound is stable. Now in the case of
the NCSYM, unfortunately the topological argument seems not to be
valid due to the high complexity of the $*$-product. Even
with this complexity, we can argue a similar mass bound as in the
following \cite{Hashimoto:1999}. 

The energy of the system without the electric field is given by
\begin{eqnarray}
  E= \tr\int\! d^3x 
  \left[
    \frac{1}{2}F_{ij}*F_{ij} +  D_i\Phi *D_i\Phi
  \right],
\label{energy}
\end{eqnarray}
where we have defined the field strength and the covariant derivative
as
\begin{eqnarray}
  F_{ij}\hs{-16pt} 
&&\equiv \p_i A_j - \p_j A_i -i A_i * A_j + i A_j * A_i,\\
D_i \Phi\hs{-16pt} 
&&\equiv \p_i \Phi - iA_i * \Phi + i \Phi * A_i.
\end{eqnarray}
We have put the Yang-Mills coupling constant $g_{\rm YM}$ equal to
unity. The $*$-product is defined with the non-commutativity parameter 
$\theta_{ij}$ by
\begin{eqnarray}
  (f*g)(x)\hs{-16pt}  
&&\equiv f(x) \exp
  \left(\frac{i}{2}\theta_{ij}
  \overleftarrow{\p_i}\overrightarrow{\p_j}\right)g(x)
     = f(x) g(x) + \frac{i}{2}\PB{f}{g}(x) + O(\theta^2),
\end{eqnarray}
where $\PB{f}{g}$ is the Poisson bracket,
\begin{eqnarray}
\PB{f}{g}(x)\equiv\theta_{ij}\p_i f(x)\,\p_j g(x) .
\end{eqnarray}
We shall take as the gauge group the simplest one $U(2)$.
Note that the group $SU(2)$ is not allowed here
since the algebra of any special unitary group is not closed when the
multiplication is defined by the $*$-product. The energy
(\ref{energy}) is bounded below by a surface integral as
follows:\footnote{
In deriving the inequality (\ref{below}), we assumed that 
$F_{ij}\pm\epsilon_{ijk}D_k\Phi$ decay sufficiently fast at the
infinity so that we can apply the formula 
$\int\!d^3x\,(f*g)(x)=\int\!d^3xf(x)g(x)$ for these quantities.
}
\begin{eqnarray}
  E\hs{-14pt}&&= 
\Half\tr\int\! d^3x 
  \Bigl[\mp\epsilon_{ijk}
    \left(
      F_{ij}*D_k\Phi + D_k\Phi * F_{ij}
    \right)
+(F_{ij}\pm\epsilon_{ijk}D_k\Phi)*(F_{ij}\pm\epsilon_{ijl}D_l\Phi)
  \Bigr]
\nn\\
&& \geq 
\tr\int\! d^3x 
  \p_k\Bigl[\mp\epsilon_{ijk}
    \left(
      F_{ij}*\Phi
    \right)
  \Bigr].
\label{below}
\end{eqnarray}
If the solution of the non-commutative BPS (NCBPS) equation,
\begin{eqnarray}
  D_i \Phi + \Half\epsilon_{ijk} F_{jk} =0 ,
\label{BPS}
\end{eqnarray}
has the same asymptotic behavior as the ordinary BPS solution, then
the energy remains the same. This fact will be confirmed to the
first non-trivial order in $\theta$ by explicitly solving the NCBPS
equation (\ref{BPS}). 

We shall solve the NCBPS equation (\ref{BPS}) to $O(\theta)$ in the
small $\theta$ expansion. Let us express the gauge field as
\begin{eqnarray}
A_i \equiv \left(A^a_i +a^a_i\right)\Half \sigma_a
+ \left(A^0_i +a^0_i\right) \Half \unit,
\label{express}
\end{eqnarray}
where the upper (lower) case component fields are of $O(\theta^0)$
($O(\theta^1)$). The scalar $\Phi$ is expressed in a similar
manner. First, the $O(\theta^0)$ part of the NCBPS equation
(\ref{BPS}) is nothing but the ordinary BPS equation, and we adopt the 
well-known BPS monopole solution as the $O(\theta^0)$ part of
the solution: 
\begin{eqnarray}
\Phi^a = \frac{x^a}{r}F(r),\quad 
A_i^a = \epsilon_{aij}\frac{x_j}{r} W(r),\quad 
\Phi^0=A^0_i=0 ,
\label{pssol}
\end{eqnarray}
where the the functions appearing in the solution are
\begin{eqnarray}
  F(r) \equiv C \coth(Cr)-\frac1r,\qquad 
  W(r) \equiv \frac1r-\frac{C}{\sinh(Cr)}.
\end{eqnarray}
The dimensionful parameter $C$ determines the mass scale of the
monopole. For later convenience, we present the asymptotic
behavior of these functions: 
\begin{eqnarray}
  F(r) = C -\frac1r + O(e^{-Cr}),\qquad 
  W(r) = \frac1r +  O(e^{-Cr}).
\label{asym}
\end{eqnarray}

Next let us proceed to the $O(\theta)$ part of the NCBPS
equation (\ref{BPS}). Plugging (\ref{express}) into the NCBPS equation 
(\ref{BPS}) and taking the $O(\theta)$ part, the $U(1)$ component
reads
\begin{eqnarray}
\p_i\varphi^0 +\Half\PB{A^a_i}{\Phi^a}+
\epsilon_{ijk}\left(\p_j a^0_k +\frac{1}{4}\PB{A^a_j}{A^a_k}\right)
=0 ,
\label{BPS2}
\end{eqnarray}
and does not contain the $SU(2)$ fields  $(a^a_i,\varphi^a)$.
On the other hand, the $SU(2)$ component of the $O(\theta)$ part of
the NCBPS equation decouples from the $U(1)$ fields, and is in fact
the linearized equation for the fluctuation $(a^a_i,\varphi^a)$
obtained from the ordinary BPS equation. Since any solution for 
$(a^a_i,\varphi^a)$ corresponds to a $\theta$-dependent gauge
transformation on the BPS solution (\ref{pssol}), we take
$a^a_i=\varphi^a=0$ hereafter.

Now our task is to solve the equation (\ref{BPS2}).
The ansatz for the BPS monopole solution (\ref{pssol}) was the
covariance under the rotation of the diagonal $SO(3)$ subgroup of
$SO(3)_{\rm gauge} \times SO(3)_{\rm space}$. In order to solve eq.\ 
(\ref{BPS2}), we put the following ansatz for the $U(1)$
fields $(a^0_i,\varphi^0)$ respecting the covariance under the
generalized rotation, in which we rotate also the parameter
$\theta_{ij}$:\footnote{
The generalized rotational covariance for $a_i^0$ allows two other
terms with different structures, $\epsilon_{ijk}\theta_{jk}$ and
$x_i\epsilon_{jkl}\theta_{jk}x_l$. However, eq.\ (\ref{BPS2}) implies
the vanishing of these two terms.
}
\begin{eqnarray}
a^0_i = \theta_{ij}x_j A(r),
\qquad   \varphi^0 = \theta_{ij} \epsilon_{ijk} x_k B(r),
\label{ansu1}
\end{eqnarray}
where $A(r)$ and $B(r)$ are functions of $r$ to be determined. 
Putting these ansatz (\ref{ansu1}) and the explicit forms of the BPS
solution (\ref{pssol}) into the differential equation (\ref{BPS2}), we 
obtain the following two equations as the coefficients of different
tensor structures:
\begin{eqnarray}
&&-A+B + rB' + \frac1{4r^2}W(W + 2F) =0, \label{acrc}\\
&&A' + 2B' - \frac{d}{d r}
  \left[\frac1{4r^2}W(W+2F) \right]=0.
\label{dif}
\end{eqnarray}
The solution to eqs.\ (\ref{acrc}) and (\ref{dif}) is given by
\begin{eqnarray}
A(r) = \frac1{4r^2}W(W + 2F) -2\frac{c_1}{r^3}+c_2 ,
\qquad
B(r) = \frac{c_1}{r^3} + c_2 ,
\label{AB}
\end{eqnarray}
with two arbitrary constant parameters, $c_1$ and $c_2$.
The parts in (\ref{AB}) containing these constant parameters are
actually solution to the homogeneous part of eq.\ (\ref{BPS2}):
\begin{eqnarray}
\p_i \varphi^0 + \epsilon_{ijk} \p_j a^0_k  =0.
\label{bpszero}
\end{eqnarray}
Since the $c_2$ part of the scalar $\varphi^0$ diverges at the
infinity, we put $c_2=0$.
As for the $c_1$ part, a careful substitution into the left hand side
of eq.\ (\ref{bpszero}) gives in fact a term proportional to
$\p_i\p_i (1/r)=-4\pi\delta^3(\rr)$.
Hence the $c_1$ part is not a solution at the origin, and we shall
also put $c_1=0$.
Finally the desired solution of the equation
(\ref{BPS2}) is
\begin{eqnarray}
a^0_i = \theta_{ij}x_j \frac1{4r^2}W(W+2F),
\qquad   \varphi^0 = 0.
\label{solution}
\end{eqnarray}
Note in particular that the scalar field receives no correction to
this order. Since the whole solution has the same leading asymptotic
behavior as the BPS solution (\ref{pssol}), we find that the
non-commutativity does not change the mass of the monopole.


\section{Non-commutative eigenvalue equation}

The configuration of the D-string suspended between two parallel
D3-branes is described by the deformation of the surface of the
D3-branes in the spirit of the BIon (Born-Infeld soliton) physics
\cite{Callan:1997,Hashimoto:1998px}. The extent of this deformation of
the D3-brane surface is given by the eigenvalues of the scalar field
on the D3-branes. We saw in the previous section that there is
no additional contribution of $O(\theta)$ in the scalar field
configuration. However, since we are now dealing with the theory with
the $*$-product, the eigenvalue problem should be different from that
in the usual commutative case. In this section, we see that the 
$O(\theta)$ terms are actually generated in the eigenvalues of the
scalar field. 

We propose that the eigenvalue equation for a hermitian matrix valued
function $M$ to be considered in the non-commutative case is 
\begin{eqnarray}
M*\vv = \lambda * \vv ,
\label{eigeneq}
\end{eqnarray}
where $\vv$ and $\lambda$ are the eigenvector and the eigenvalue,
respectively. Though there are other candidates for the non-commutative
eigenvalue equation, the present one (\ref{eigeneq}) has advantages
over the others in various respects as we shall see in this and the
final sections.

For solving (\ref{eigeneq}) to $O(\theta)$, let us make the expansion
\begin{eqnarray}
  M=M_0 + M_1, \quad
  \vv = \vv_0 + \vv_1, \quad
  \lambda = \lambda_0 + \lambda_1 ,
\label{0+1}
\end{eqnarray}
where the subscript number specifies the order of $\theta$.
Then, the $O(\theta^0)$ part of (\ref{eigeneq}) is
$M_0\vv_0=\lambda_0\vv_0$, and the $O(\theta)$ part reads
\begin{eqnarray}
M_0 \vv_1 +M_1 \vv_0+ \frac{i}{2}\PB{M_0}{\vv_0}
= \lambda_0 \vv_1 + \lambda_1 \vv_0 
+  \frac{i}{2}\PB{\lambda_0}{\vv_0} .
\end{eqnarray}
Multiplying $\vv_0^\dagger$ from the left, we obtain the formula which
gives the $O(\theta)$ part of the eigenvalue:
\begin{eqnarray}
\lambda_1 = \frac{1}{\vv_0^\dagger \vv_0}\left(
\frac{i}{2}
\vv_0^\dagger\PB{M_0-\lambda_0\unit}{\vv_0}
+ \vv_0^\dagger M_1\vv_0
\right) .
\label{formula_general}
\end{eqnarray}
In view of the application to the present NCBPS solution, let us
consider the particular case with
\begin{eqnarray}
M_0= m_0(\rr)\,\hx_a\sigma_a\quad (\hx_i\equiv x_i/r),
\qquad
M_1=0 ,
\end{eqnarray}
and hence $\lambda_0=\pm m_0(\rr)$ and $\hx_a\sigma_a\vv_0=\pm\vv_0$.
Then, eq.\ (\ref{formula_general}) is calculated to give
\begin{eqnarray}
\lambda_1 = \frac{i}{2\vv_0^\dagger \vv_0}\frac{m_0(\rr)}{r}
\theta_{ij}\left(
    \vv_0^\dagger \sigma_i \p_j \vv_0 
    \mp \hx_i\vv_0^\dagger \p_j \vv_0 
  \right)
=-\frac{m_0(\rr)}{2r^2}\theta_i\hx_i ,
\label{formula}
\end{eqnarray}
with $\theta_i\equiv(1/2)\epsilon_{ijk}\theta_{jk}$.
Note that $\lambda_1$ (\ref{formula}) is independent of the sign
of $\lambda_0$.
We obtained the last expression of (\ref{formula}) using the explicit
form $\vv_0=(x_1 - ix_2,\pm r - x_3)^{\rm T}$.
However, the general formula for $\lambda_1$, eq.\
(\ref{formula_general}), is in fact invariant under the local 
phase and scale transformation of $\vv_0$ due to the identity
$\vv_0^\dagger\PB{M_0-\lambda_0\unit}{f\unit}\vv_0=0$ valid for any
$f(\rr)$.
This corresponds to the fact that the eigenvalue $\lambda$ of
(\ref{eigeneq}) is invariant under the right multiplication $\vv\to\vv
*h$ for an arbitrary $h(\rr)$.
We shall discuss the gauge transformation property of the eigenvalues
in the final section.

Let us evaluate various eigenvalues of the system using the formula
(\ref{formula}). First, the scalar eigenvalues are obtained by
substituting $m_0(\rr)= F(r)/2$ as
\begin{eqnarray}
  \lambda_\Phi = \pm \Half F(r) - \frac{\theta_i \hx_i}{4r^2} F(r) +
  O(\theta^2). 
\label{higgsev}
\end{eqnarray}
Next, we shall consider the eigenvalues of the magnetic field
$B_i\equiv(1/2)\epsilon_{ijk}F_{jk}$ near the infinity $r\to\infty$.
The magnetic field itself is given from the solution (\ref{solution})
as\footnote{
Since the definition of the magnetic field contains the $*$-product in
itself, we should calculate also the Poisson bracket term. However, 
this term contributes only to the $O(1/r^4)$ part in (\ref{mag}).}
\begin{eqnarray}
  B_i = -\frac{\hx_i}{2r^2}\hx_a \sigma_a + 
\frac{C}{2r^3}
\left(\delta_{ij}-3\hx_i\hx_j \right)\theta_j\unit
+O\left( \frac1{r^4}\right).
\label{mag}
\end{eqnarray}
We would like to evaluate the $O(\theta)$ contribution to the
eigenvalues by putting $m_0(\rr)=-\hx_i/2r^2$ and
$M_1(\rr,\theta)=
C\left(\delta_{ij}-3\hx_i\hx_j \right)\theta_j\unit/2r^3$.
Since $m_0(\rr)$ in this case is $O(1/r^2)$, using
the formula (\ref{formula}), the order of the correction to the 
eigenvalues from this part is found as $O(1/r^4)$. Thus near the
infinity, the $O(1/r^3)$ part of the eigenvalues of the magnetic field
is saturated by $m_1$.


\section{Interpretation of the eigenvalues}

In this section, we shall see how the eigenvalues (\ref{higgsev}) and
(\ref{mag}) reproduce the brane configuration depicted in fig.\
\ref{fig1}.  
In the brane picture, the end of the D-string is seen as a magnetic
charge in a single D3-brane worldvolume theory. The prediction from
the brane configuration is that the magnetic charge on each end of the 
D-string is actually moved in different directions between the upper
and lower D3-branes, as shown in fig.\ \ref{fig1}.
This shift is specified by the spatial vector $\delta_i$. 

Now, the eigenvalues of the magnetic field (\ref{mag}) indicate that
the $U(1)$ part of the magnetic field exhibits a dipole
structure. This structure is exactly the one expected from the brane
picture above. Noting that the zero-th order solution (\ref{pssol})
represents $-1/2$ charge on the upper D3-brane and $1/2$ charge on the 
lower, it is easy to derive the non-locality $\delta_i$ as
\begin{eqnarray}
  \delta_i = C \theta_i.
\label{magres}
\end{eqnarray}
This result verifies the prediction of ref.\ \cite{Hashimoto:1999}
with the identification $C=U$.

\begin{figure}[htb]
\begin{center}
\parbox[b]{130mm}{
\begin{center}
\leavevmode
\epsfxsize=120mm
\put(330,120){{\Large $x_1$}}
\put(170,224){{\Large $\lambda_\Phi$}}
\epsfbox{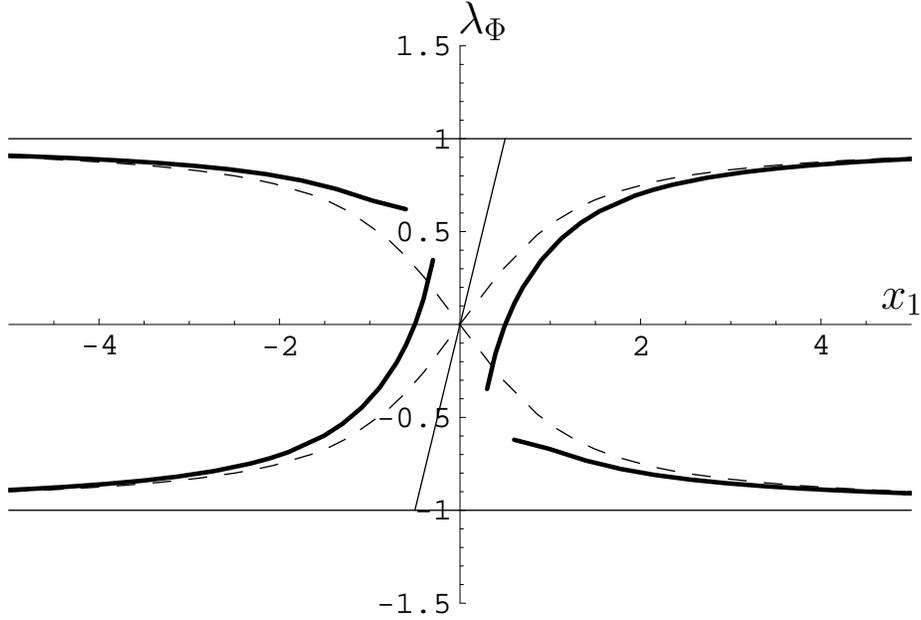}
\caption{The eigenvalues of the solution for the scalar
  field. We choose $C=2$ and $\theta_{23}=1/2,$ and other $\theta$'s
  are set to zero. }
\label{fig2}
\end{center}
}
\end{center}
\end{figure}

Although the magnetic charges are expected to indicate only the
locations of the ends of the D-string, the eigenvalues of the scalar
field must reproduce not only the locations of the ends but also the
slope of the D-string. In fact, the asymptotic behavior of the
eigenvalues (\ref{higgsev}) is given using (\ref{asym}) as
\begin{eqnarray}
&&\lambda_\Phi= \pm \frac{C}{2} \mp \frac{1}{2r}
+\left(-\frac{C}{4r^2}+\frac{1}{4r^3}\right)\theta_i \hx_i
+ O(e^{-Cr})
\nn\\
&&\phantom{\lambda_\Phi}
=\pm \frac{C}{2}\mp\frac{1}{2}\left|
x_i\mp\left(\frac{C}{2}-\frac{1}{2r}\right)\theta_i\right|^{-1}
+ O(e^{-Cr}) .
\label{aslam}
\end{eqnarray}
Eq.\ (\ref{aslam}) implies first that in the upper and the lower
D3-brane the end of the D-string sits at $x_i = C \theta_i/2$ and
$x_i = -C \theta_i/2$, respectively.
Hence the non-locality is precisely
given by $\delta_i = C \theta_i$, in agreement with the result
(\ref{magres}) from the magnetic field.
Secondly, in order to read off the slope of the D-string from
(\ref{aslam}), we rewrite it as
\begin{eqnarray}
&&\lambda_\Phi= \pm \frac{C}{2}\mp\frac{1}{2}\left|
x_i-\lambda_\Phi\theta_i\right|^{-1}
+ O(e^{-Cr}) .
\label{aslam2}
\end{eqnarray}
This equation means that, for a given value of $\lambda_\Phi$, the
corresponding worldvolume coordinate is located on a sphere with
its center at $x_i=\lambda_\Phi\theta_i$.
Interpreting the trajectory of the center as the D-string, 
our analysis reproduces precisely the tilt of the suspended D-string.

In fig.\ \ref{fig2} we present the curves of the eigenvalues of the
scalar field. 
The thin straight lines represent the brane configuration of fig.\
\ref{fig1}. The dashed curves denote the eigenvalues of the scalar
field with $\theta=0$. Comparing these with the bold curves
representing the eigenvalues (\ref{higgsev}) with $\theta \neq 0$,
we can read off the simple brane configuration of ref.\
\cite{Hashimoto:1999}.
(The reason why the bold curves are cutoff for small $|x_1|$ in
fig.\ \ref{fig2} is that the $O(\theta)$ term of the scalar
eigenvalues (\ref{higgsev}) are actually divergent at the origin
$r=0$. We shall discuss this problem in the next section.)


\section{Summary and discussion}

In this paper, we solved the BPS equation of the NCSYM to the
first non-trivial order in $\theta$. Evaluating the eigenvalues of the 
solution, we explicitly showed that the solution exhibits the brane
configuration of the tilted D-string, given in ref.\
\cite{Hashimoto:1999}. Magnetic field has the dipole structure, and
the scalar field is also shifted to reproduce the tilted trajectory 
of the D-string.

Some comments are in order. 
Our first comment is on the gauge transformation property of the
eigenvalue $\lambda$ in our non-commutative eigenvalue equation
(\ref{eigeneq}).
Of course, the eigenvalue $\lambda$ in eq.\ (\ref{eigeneq}) is never
strictly invariant under the local gauge transformation of $M$,
\begin{equation}
M\to U^{-1}*M*U ,
\label{eq:gtrM}
\end{equation}
where $U^{-1}$ is the inverse of $U$ with respect to the $*$-product,
$U*U^{-1}=U^{-1}*U=\unit$.
However, we can show that the eigenvalue has a fairly nice property
under (\ref{eq:gtrM}).
Consider an infinitesimal gauge transformation $\de$ on $M$ with
$U=\unit + i\epsilon$ ($\epsilon^\dagger =\epsilon$), 
\begin{equation}
\de M=i(M*\epsilon-\epsilon*M) .
\end{equation}
Letting $\de$ act on (\ref{eigeneq}) and $*$-multiplying the resultant
equation by $\vv^\dagger$ from the left, we obtain
\begin{eqnarray}
\vv^\dagger *\de\lambda *\vv = i\vv^\dagger
*(\lambda *\epsilon - \epsilon *\lambda)*\vv .
\label{benri}
\end{eqnarray}
Taking the $O(\theta)$ part of (\ref{benri}) and using
$\de\lambda_0=0$, $\de\lambda_1$ is given as
\begin{equation}
\de\lambda_1 = \vv_0^\dagger\PB{\epsilon}{\lambda_0}\vv_0 ,
\label{eq:delambda1}
\end{equation}
for a normalized $\vv_0$.
Eq.\ (\ref{eq:delambda1})
implies that, at least to $O(\theta)$, the gauge transformation
corresponds to a coordinate transformation on the eigenvalue
$\lambda(\rr)$. In the $U(2)$ case with 
$\epsilon(\rr)=\epsilon_a(\rr)\sigma_a + \epsilon_0(\rr)\unit$,
the form of the coordinate transformation is
$\de x_i = -\theta_{ij}\left(
\p_j\epsilon_0 + \vv_0^\dagger\sigma_a\vv_0\,\p_j\epsilon_a
\right)$.

Therefore, we have shown that the eigenvalue of (\ref{eigeneq}) for
$M$ and the one for $U^{-1}*M*U$ are related by a
coordinate transformation on the D3-branes and hence are physically
equivalent, at least to the first non-trivial order in $\theta$.
Recalling also that the eigenvalue of (\ref{eigeneq}) is independent 
of the choice of the eigenvector and that the scalar eigenvalue
(\ref{higgsev}) has an invariance under the simultaneous rotation of
$x_i$ and $\theta_{ij}$, the eigenvalue equation (\ref{eigeneq}) we
have adopted is a satisfactory one.
These good properties, in particular,
the gauge transformation property of $\lambda$, are expected to
persist to higher orders in $\theta$.

Our second comment is on another type of non-commutative eigenvalue
equation,
\begin{eqnarray}
M*\vv =  \vv  * \lambda ,
\label{gyaku}
\end{eqnarray}
where, compared with (\ref{eigeneq}), $\vv$ and $\lambda$ are
interchanged on the right hand side.
Eq.\ (\ref{gyaku}) has a property that the eigenvalue $\lambda$ is
invariant under the gauge transformation (\ref{eq:gtrM}) of $M$.
However, for a given $M$ the eigenvalues are not unique and do depend
on the choice of the eigenvectors $\vv$. What is worse, 
in the analysis similar to those in Sec.\ 3 by adopting 
(\ref{gyaku}), we can show that it is impossible to obtain the
the scalar eigenvalues possessing an invariance under the simultaneous
rotation of $x_i$ and $\theta_{ij}$.
These are the reasons why we did not adopt (\ref{gyaku}). 

Our final comment is concerning the singular nature of the scalar
eigenvalues (\ref{higgsev}) at the origin $r=0$.
Since we have adopted the eigenvalue equation (\ref{eigeneq}), the
matrix whose diagonal entries are these eigenvalues is no longer a
solution of the BPS equation. Hence although the eigenvalues
(\ref{aslam}) diverge at the origin, the energy of the configuration
is still finite. We would need a technique beyond the expansion in
powers of $\theta$ to dissolve the singularity at the origin.

Our analysis in this paper can straightforwardly be applied to the case
of dyons. We obtain a result consistent with the non-locality
predicted by \cite{Hashimoto:1999}. It would be an interesting subject 
to study the non-commutative 1/4 BPS dyons using our formulation.


\section*{Note added} 
\vs{-5mm}
While writing this paper, we became aware of the paper \cite{Bak:1999}
which has an overlap concerning the solution of Sec.\ 2.
   

\section*{Acknowledgments}
\vs{-5mm}
We would like to thank A.\ Hashimoto, T.\ Hirayama, S.\ Imai, H.\
Kawai, I.\ Kishimoto, T.\ Kugo, Y.\ Okawa and S.\ Terashima for
valuable discussions and comments. This work is supported in part by
Grant-in-Aid for Scientific Research from Ministry of Education,
Science, Sports and Culture of Japan (\#3160, \#09640346, \#4633).
The work of K.\ H.\ and S.\ M.\ is supported in 
part by the Japan Society for the Promotion of Science under the
Predoctoral Research Program.


\begingroup\raggedright\endgroup

\end{document}